\documentclass[aps,pra, twocolumn,nofootinbib]{revtex4-1}
\usepackage[utf8]{inputenc}
\usepackage{amsmath}
\usepackage{braket}
\usepackage[bottom]{footmisc}
\usepackage{appendix}
\newtheorem{theorem}{Theorem}
\newtheorem{definition}{Definition}
\newtheorem{corollary}{Corollary}

\newcommand{\proj}[1]{| #1 \rangle \langle #1 |}
\newcommand{\op}[2]{| #1 \rangle \langle #2 |}
\newcommand{\tr}[1]{\mathrm{Tr} \left( #1  \right) }
\newcommand{\trs}[2]{\mathrm{Tr}_{#2} \left( #1  \right) }
\newcommand{\identity}{I}

\begin{document}

\title{Channels, measurements and post-selection in quantum thermodynamics}
\author{Tom Purves}\email{tom.purves@bristol.ac.uk} \affiliation{H.H. Wills Physics Laboratory, University of Bristol, Tyndall Avenue, Bristol, BS8 1TL, U.K.}

\author{Anthony J. Short}\email{tony.short@bristol.ac.uk} \affiliation{H.H. Wills Physics Laboratory, University of Bristol, Tyndall Avenue, Bristol, BS8 1TL, U.K.}

\date{August 2019}

\begin{abstract}
We analyse the benefit, in terms of extracting work, of having a single use of a  quantum channel or measurement in quantum thermodynamics. This highlights a connection between unital and catalytic channels, and  some subtleties concerning the conditional work cost of implementing a measurement given that a certain result was obtained. We also consider post-selected measurements, and show that any non-trivial post-selection leads to an unbounded work benefit.  
\end{abstract}

\maketitle
\section{Introduction}

Thermodynamics is perhaps the only known theory of physics that never comes up short. It applies ubiquitously across many situations well outside of its original realm of conception - to understand macroscopic thermal machines such as steam engines - and holds when considering both tiny objects such as collections of few atoms \cite{Toyabe2015}, and massive exotic phenomena such as black holes \cite{Carlip2014}. Quantum thermodynamics has received a renewed interest recently, with many interesting results already having been obtained. These include thermodynamics at the nano-scale \cite{Alicki1979, Allahverdyan2004,Allahverdyan2008, Linden2010, Allahverdyan2011, Horodecki2013, Anders2013, Skrzypczyk2014, Richens2016}, thermal machines that operate in these regimes \cite{Scovil1959, Scovil1967, Tonner2005, Malabarba2015, Kosloff2014}, equilibration \cite{Reimann2008, Short2009, Linden2009, Gogolin2016},  results pertaining to the relationship between thermodynamic resources \cite{ Guryanova2016,  Aberg2014, Horodecki2013_2}, Landauer erasure \cite{Lloyd1997, Esposito2011, Vaccaro2011}, theoretical results concerning the second law of thermodynamics \cite{Janzing2000, Mahler2005, Verley2012, Brunner2012, Brandao2015} as well as many interesting information theoretic results \cite{Rio2011, Dahlsten2011, Egloff2012}. For a topical review, the authors recommend \cite{Goold2016}.

One of the first questions that thermodynamics addressed is the maximal amount of work that can be extracted from a thermodynamic transformation at a particular temperature? For classical physics, this is given by the reduction in \textit{free energy} of the system under consideration. Given that thermodynamics typically concerns large systems, one might expect this result to hold only for the collective processing of many copies of a quantum system. However, 
an analogous result can be re-derived for \textit{individual} quantum systems \cite{Skrzypczyk2014}, when work is understood as the average energy change of a quantum `weight on a string' (a quantum system that can be raised or lowered continuously to store energy) .  

The framework introduced in \cite{Skrzypczyk2014} originally imposed only average energy conservation on the system, thermal bath, and weight, but was adapted in \cite{Malabarba2015} to include strict energy conservation, which we will use here. These papers considered protocols in which the system, bath and weight undergo unitary transformations. In this paper, we will  extend these results to include general quantum channels and measurements. In particular, we consider the work benefit of having access to a single use of a quantum channel or measurement, as well the thermodynamic power of post-selection.  

A quantum channel describes a unitary evolution in which a system interacts with an unobserved ancilla. During quantum computation or communication, we typically seek to minimize interactions of this type, to avoid decoherence. From a thermodynamic standpoint, however, we show that any non-unital channel leads to a work benefit. In the context of the second law of thermodynamics, these work benefits can be seen as a consequence of a Maxwell-demon type effect \cite{Bennett1987} due to discounting the increasing entropy of the device. We also explore the connection with catalytic channels, in which the state of the ancilla remains unchanged.

Other recent work in this area has taken a resource-theoretic approach to the thermodynamics of quantum channels, studying the asymptotic limit of many uses of the channels. This has yielded the asymptotically extractable work for a set of channels \cite{Navascu2015}, as well as a work related notion of simulability of one channel by another \cite{Faist2019}. A key difference with  our approach is that we only consider a single use of the channel. The work cost of implementing a quantum channel has also been studied in an alternative framework \cite{faist2015}.

We also quantify the work benefit of a general quantum measurement, and highlight some interesting subtleties concerning the conditional work benefit of measurement given that a certain result was obtained. Recently the energy cost of measurement \cite{Abdelkhalek2016}, and the work loss of erasing quantum coherences via measurement \cite{Kammerlander2016} has been considered, as well as the use of quantum measurements as a cooling engine \cite{Buffoni2019}.

Finally, we analyse the thermodynamics of post-selected measurements, where we only consider cases in which some subset of the outcomes occurs. We show that post-selection is extraordinarily powerful in a thermodynamic setting. In particular, it is possible to use any (non-trivial) post-selected measurement to extract unbounded amounts of work, even with a single use of the measurement. 

The remainder of the paper is structured as follows. In section \ref{sec::Framework} we lay out the thermodynamic framework for our analysis. In section \ref{sec::channels} we consider the work benefit of quantum channels, and in section \ref{sec::measurements} we consider quantum measurements. Then in section \ref{sec::post} we consider post-selection, followed by conclusions in section \ref{sec::conclusions}.

\section{Framework} \label{sec::Framework}

We begin by outlining a simple framework for quantum thermodynamics, in which  internal energy, heat and work are understood in terms of average energies \cite{Skrzypczyk2014, Malabarba2015}. This yields very similar results to the classical case, and allows one to prove the thermodynamic laws. Consider a system $S$ with Hamiltonian $H_s$ and initial state $\rho_s$, interacting with a thermal bath at temperature $T$. The internal energy of the system corresponds to its average energy  $U =E_s = \texttt{Tr}[\rho_s H_s]$, and its entropy is given by the Von Neumann entropy $S(\rho_s)=-\texttt{Tr}[\rho_s \text{ln}\rho_s]$. As we will see, in this scenario the amount of useful work which can be extracted  is not solely determined by the system's internal energy, but rather by its free energy 
 \begin{equation}
    F_s =E_s - TS(\rho_s).
\end{equation}
To model the thermal bath, we assume that it is sufficiently large that we can find within it a quantum system $B$ with any Hamiltonian $H_b$ in a thermal state  at temperature $T>0$. The thermal state $\tau_b$ is the state which minimizes the free energy of the bath $F_b=E_b-TS (\tau_b)$, where $E_b = \tr{H_b \tau_b}$ is the average energy of the bath.  This yields
\begin{equation} 
\tau_b=\frac{e^{-\frac{H_b}{T}}}{\tr{e^{-\frac{H_b}{T}}} },
\end{equation} 
where for simplicity we have chosen  units in which Boltzmann's constant is unity, $k_B=1$. 

A thermodynamic protocol corresponds to selecting a particular thermal state from the bath, which we assume to be initially uncorrelated with the system  (i.e. $\rho=\rho_s \otimes \tau_b$), and then implementing a unitary transformation $U$ on the system and bath $\rho'={U}\rho {U}^\dagger$. During any such transformation, the internal energy of the system changes as 
\begin{equation}
    \Delta U = \Delta E_s = \texttt{Tr}[H_s\rho_s'-H_s\rho_s].
\end{equation}
Similarly, we define the heat flow  as the change in the average energy of the bath (with the convention that positive heat flows correspond to energy flowing into the bath) 
\begin{equation}
    Q = \Delta E_b =  \texttt{Tr}[H_b\rho_b'-H_b\tau_b]. 
\end{equation}
The work $W$ is then defined implicitly by the first law of thermodynamics 
\begin{equation}\label{eq::firstlaw}
    \Delta U +  Q +  W = 0 
\end{equation}
This corresponds to the case in which batteries are treated \emph{implicitly}. We extend these results to incorporate an \textit{explicit} battery, and thereby reinforce their relevance, in the next subsection. 

To prove the second law, we first consider the entropy changes which occur during a protocol. Due to our assumption that the system and bath are initially uncorrelated we  have $S(\rho)=S(\rho_s \otimes \tau_b)=S(\rho_s)+S(\tau_b)$. Using the fact that the total entropy is conserved under unitary evolution, and that the entropy is subadditive, we find
\begin{equation}\label{eq::entropy}
    S(\rho_s)+S(\tau_b) = S(\rho)=S(\rho') \leq S(\rho_s')+S(\rho_b')
\end{equation}
where $\rho_s' = \trs{\rho'}{b}$ is the final reduced state of the system, and $\rho_b'= \trs{\rho'}{s}$ is the final reduced state of the bath. Hence 
\begin{equation} \label{eq:entropyincrease} 
\Delta S_s + \Delta S_b \geq 0
\end{equation} 
Combining this with the first law, we obtain 
\begin{equation} 
\Delta F_s + \Delta F_b + W \leq 0
\end{equation} 
As the thermal state is defined to be the state of minimal free energy we must have $\Delta F_b \geq 0$, and hence 
\begin{equation} \label{eq:work_extraction}
    W \leq - \Delta F_s .
\end{equation}
Hence, as in standard classical thermodynamics, if the system undergoes any cyclic process in which it begins and ends in the same state, then $\Delta F_s=0$ and thus $W \leq 0$  (no net work can be extracted). This is a statement of the second law. 

Furthermore, for any desired transformation of the system from $\rho_s$ to $\rho'_s$, there exists a protocol which comes arbitrarily close to extracting the optimal work given by \eqref{eq:work_extraction} and produces a final state arbitrarily close to $\rho'_s$ \footnote{Note that when $\rho'_s$ is full rank, we can generate it exactly whilst extracting work arbitrarily close to optimal.} \cite{Skrzypczyk2014}. First, consider the case in which $\rho_s$ and $\rho'_s$ are diagonal in the energy basis, and denote them by $\rho_{0}$ and $\rho_{N+1}$ respectively. Next, find a sequence of $N$ thermal subsystems in the bath with states $\rho_{1} \ldots \rho_{N}$, such that $\rho_{k}$ is $\epsilon$-close to $\rho_{k-1}$ for all $k$. Then,in the $k^{th}$ step of the protocol, we swap the current state of the system with the state $\rho_{k}$ in the bath. After $N$ steps, we will have achieved the desired final state. As each step in this protocol is a swap operation which does not introduce correlations, it satisfies  $\Delta S_s + \Delta S_b = 0$, and hence $\Delta F_s + \Delta F_b + W = 0$. Furthermore, as each subsystem in the bath is initially at a minimum of free energy, and is transformed into a new state which is $\epsilon$-close to its original state, we find  $ \Delta F_b = O(\epsilon^2)$ during each step. After $N= O(\frac{1}{\epsilon})$ steps we therefore find that $W = -\Delta F_s - O(\epsilon)$, which can be made arbitrarily close to the bound given in \eqref{eq:work_extraction} by choosing $\epsilon$ sufficiently small. If $\rho_s$ and $\rho'_s$ are not diagonal in the energy basis, we can perform an additional unitary rotation on the system at the beginning and end of the protocol to transform its eigenbasis. Such unitary transformations satisfy $W = - \Delta F_s$ and therefore do not lead to any loss of work. 

A version of the third law of thermodynamics, that no finite process can transform a finite temperature state into a zero temperature state, can also be seen by noting that these states are operators of different rank, and that a unitary transformation preserves the rank. For a more detailed analysis of this, see \cite{masanes2017}.

The two main assumptions of this framework are that the initial state is a product state and that the evolution is unitary. The former is designed to avoid conspiratorial situations in which the history of the bath and system is complex. Introducing correlations between initial states of the system and bath allows one to smuggle in thermodynamic resources from the outset \cite{PerarnauLlobet2015}, and can be analysed within this framework by considering all of the correlated subsystems as forming a larger non-thermal system. The assumption of unitarity prevents cheating by using ancillas as additional thermodynamic resources which are not accounted for. However, an aim of this paper is to explore in more detail how  non-unitary transformations such as quantum channels and measurements fit into this framework, and to quantify the additional power they bring. 

A natural criticism of the implicit battery framework above is that it lacks a physical mechanism for work extraction, and does not specify how the work is stored for later use (for example, by raising a weight on a string). Furthermore, it only incorporates conservation of average energy (via \eqref{eq::firstlaw}) rather than of the total energy distribution. One might wonder if a more physical model would place additional limitations on the extractable work. To address these issues, we next introduce an \textit{explicit} battery into the framework.  

\subsection{Explicit batteries} \label{sec::explicit}

For our purposes, a battery is a device to store the work generated during a process, or to provide work to implement it. Although typical batteries rely on chemical energy, it is  convenient here to consider an idealised mechanical battery corresponding to a `weight on a string', which can be raised or lowered during a process to store or extract energy. 

The framework incorporating an explicit battery is largely the same as above, except that we now introduce a third system, which we refer to as the weight. We model the weight as a quantum system with one continuous spatial degree of freedom $\hat{x}$ (it's `height'), and Hamiltonian $H_w=mg \hat{x}$. For simplicity, we choose the mass such that $mg = 1 J m^{-1}$, so that we can use distances and energies interchangeably. Now that we have included a battery explicitly, we impose strict energy conservation on the combined system, bath and weight, such that the allowed unitary transformations $U$ satisfy 
\begin{equation} \label{eq:total_energy}
    [U,H_s+H_b+H_w]=0.
\end{equation}
This ensures that no additional energy can be used other than that provided by the battery \footnote{Note that this differs from \cite{Skrzypczyk2014}, in which only average energy is presumed, but follows instead the revised framework presented in \cite{Malabarba2015}}. 

For the same reason as in the implicit battery case  (we want to avoid smuggling in useful correlations), we assume that all states describing the system, weight and bath begin uncorrelated, $\rho=\rho_s \otimes \tau_b \otimes \rho_w$. 

Given this setup, we now no longer need define the work implicitly, instead we define it as the change in average energy of the weight. All of the three energies appearing in the first law are then defined similarly, 
\begin{eqnarray}
         Q =& \tr{H_b\rho_b'-H_b\tau_b}\\
    \Delta U =& \tr{H_s\rho_s'-H_s\rho_s}\\
          W =& \tr{H_w\rho_w'-H_w\rho_w}
\end{eqnarray}
and the first law $\Delta U +  Q +  W = 0 $  follows straightforwardly  from \eqref{eq:total_energy}. 

Finally, we require that any allowed unitary $U$ is translationally invariant on the weight. Translation operators on the weight are given by
\begin{equation}\label{eq::translation}
    \Gamma_{E}=e^{-\frac{i}{\hbar}E\,\hat{p}_w}, 
\end{equation}
where $\hat{p}_w$ is the momentum operator on the weight space, so we require that 
\begin{equation}
    [U, \Gamma_E] =0   \; \forall\, E
\end{equation}
This prevents the weight from being used as a thermodynamic resource (for example by treating it as a cold reservior), and allows us to prove that \eqref{eq:work_extraction}, and thus the second law, hold in the explicit battery framework (see appendix \ref{app::secondlaw}). Furthermore, we can again achieve protocols which come arbitrarily close to saturating this bound if either the initial system state is diagonal in its energy basis and $U$ is a permutation of energy levels, or if the weight is prepared in a broad coherent state of energy (sharply peaked around zero in momentum space). For further details of these proofs see appendix \ref{app::secondlaw}.

Other works have introduced different models of batteries, including a two-level \emph{wit} \cite{Horodecki2013}, a weight with discrete spacing \cite{Aberg2014} or energies bounded from below \cite{Lipkabartosik2019}, which lead to additional constraints on the initial state or different results in terms of work extraction. Our motivation for considering the model above is it leads to a simple model, with the closest parallels to classical thermodynamics. The model can also be extended to incorporate an explicit quantum clock, which can implement a time-dependent protocol via a fixed time-independent Hamiltonian \cite{Malabarba2015}

The fact that we recover the same results with explicit and implicit batteries, offers an enhanced  justification for using the simple framework of implicit batteries.

\section{Channels}\label{sec::channels}

In the framework defined above, we have exclusively considered unitary protocols. Here, we expand this to include more general transformations (often referred to as quantum channels). As well as being interesting in its own right, this is also an important preliminary to considering the thermodynamics of measurements and post-selection in later sections. 

In particular, we will consider the thermodynamic advantages that channels yield (if any), above what is possible in the framework above, and the corresponding thermodynamic costs. 

Consider that you have a single use of a quantum channel $C$, which is a completely positive, trace preserving map from density operators to density operators, which can be applied to a target system $T$ with Hamiltonian $H_t$. For generality, we also allow the use of an ancillary system $A$ (with Hamiltonian $H_a$) on which the channel does not act.

The system $S$ considered in the thermodynamic framework  in the previous section is then composed of $T$ plus $A$. It will be helpful to denote the action of $C$ on the system $S$ by the channel $C_s$ (this consists of applying the channel $C$ to the target system and the identity channel to the ancilla. i.e. $C_s = (I \otimes C)$).

We  first consider the implicit battery framework, in which the channel is treated as a primitive operation. We will then show that the same results are obtained in an explicit framework in which the channel is implemented via a unitary on ancillary system, through strictly energy-conserving interactions with an explicit battery.

\subsection{Work benefit} 

We will quantify the work benefit of the channel by the maximum work that can be extracted in a cyclic process in which the channel is applied to the target system once. In addition, we allow any operations from the framework defined above (i.e. unitary operations involving a thermal bath at temperature T). By a cyclic process, we mean that the initial and final states of the system must be the same. As mentioned in section \ref{sec::Framework}, without the use of the channel such a cyclic process cannot generate net work. 

There are two natural scenarios we could consider with respect to the work cost of implementing the channel. The first is that the channel is free to perform (i.e. if it requires work to implement, this is drawn from its own internal battery). A downside of this is that even unitary operations would give an thermodynamic advantage, for example, by raising a  ground state of the system to an excited state. We discuss this case further  in appendix \ref{app::alternatequantifier}, explaining how it modifies the results obtained in the paper. 

As we are primarily interested in the thermodynamic advantage of operations outside the standard framework, we instead consider a second approach in the main paper. In particular, we assume that the channel is `plugged in' to the same battery as other operations in the framework, so that any work that is required to perform the channel (or gained by performing it) is accounted for in the same way as for unitary transformations in the framework. Nevertheless, we find that many non-unitary channels can be used to generate net work in a cyclic process.

A general protocol $P$ for work extraction involves the following steps 
\begin{enumerate} 
\item The system and bath are initialized in an uncorrelated initial state  $\rho=\rho_s \otimes \tau_b$. 
\item The channel is applied to the target system, giving the state $\sigma =  C_s[\rho_s] \otimes \tau_b$.
\item A unitary interaction is performed, transforming the state into $\sigma' = U \sigma U^{\dagger}$, where $\sigma'_s = \rho_s$ (i.e. the system returns to its initial state).
\end{enumerate} 

Specifying a protocol $P$ includes choosing the desired ancillary and bath systems to complement the target system, as well as the initial state $\rho_s$ of the system and the unitary $U$. 

The total work extracted by the protocol $P$ is given by the sum of the work gained in each step. 
\begin{align} 
W^{\mathrm{total}}_{C,P} = W_1 + W_2 + W_3 
\end{align} 

In order to calculate this, we must first define the work gained when applying the channel to the target system in step 2). As this does not involve an interaction with the thermal bath (and hence $\Delta Q=0$) it follows from \eqref{eq::firstlaw} that 
\begin{definition} \label{def::implicitworkcostC}
The work gained when applying the channel $C$ to the target system in a state $\rho_t$ (part of a larger system in state $\rho_s$)  is given by
\begin{align} 
W^{\mathrm{apply}}_C \equiv - \Delta U_{\rho_t \rightarrow C[\rho_t]} &=   \tr{ H_t \rho_t - H_t C[\rho_t] } \nonumber \\
& = \tr{H_s \rho_s - H_s C_s[\rho_s] }
\end{align} 
\end{definition} 
Using this definition and the results of the previous section, the work gain in each step can be bounded:
\begin{align} 
W_1 &=0 \\
W_2 &= W^{\mathrm{apply}}_C =  E(\rho_s) - E(\sigma_s) \\ 
W_3 &\leq F(\sigma_s) - F(\rho_s). 
\end{align} 
Hence using $F(\rho) = E(\rho) - T S(\rho)$,
\begin{align} 
W^{\mathrm{total}}_{C,P} &\leq T \left( S(\rho_s) - S(\sigma_s) \right) \nonumber  \\
 &= T \left( S(\rho_s) - S(C_s[\rho_s] \right)  \nonumber \\
&\leq T \left( S(\rho_t) - S(C[\rho_t] \right), \label{eq:worktotal}
\end{align} 
where the last step follows from the monotonicity of the mutual information under the action of local channels \footnote{|The mutual information is defined as $I(A:B) =S(A) + S(B) - S(A,B)$. It can be shown \cite{NielsenChuang} that $I(A:C[B]) \leq I(A:B)$, which implies $S(A,B) - S(A,C[B]) \leq S(B) - S(C[B]) $ }. 

Note  that for any $\rho_t$ we can consider a protocol in which $\rho_s= \rho_t$ (i.e. no ancillas are used), and $U$ is a thermodynamically efficient protocol for transforming $C[\rho_t]$ into $\rho_t$. Hence there exist protocols which come as close as desired to saturating \eqref{eq:worktotal} for any $\rho_t$. 

Finally, we optimize over protocols to find the work benefit of the Channel $C$.
\begin{theorem}\label{thrm::workbenefitC}
The  work benefit of the channel $C$ is given by 
\begin{align}
 W^{\mathrm{total}}_{C} &\equiv \sup_P\, W^{\mathrm{total}}_{C,P} = \max_{\rho_t} \, T \left( S\left(\rho_t\right) - S\right(C[\rho_t]\left) \right).
\end{align}
\end{theorem} 
Note that $W^{\mathrm{total}}_{C} \geq 0$ for all channels, as can be seen by considering the maximally mixed initial state $\rho_t =\identity/d$. However from a thermodynamic perspective, a channel is useful (i.e. $W^{\mathrm{total}}_{C}>0$) if and only if it can reduce the entropy of some state. Given this result, it is interesting to identify which channels have this property. We now prove 
\begin{corollary} \label{cor:unital} 
A channel provides no  work benefit  if and only if it is  unital.
\end{corollary} 
Unital quantum channels are those that preserve the maximally mixed quantum state, $C\left(\frac{\identity}{d}\right)=\frac{\identity}{d}$. These include unitary channels or mixtures of unitaries, but also other possibilities \cite{Ritter2005}. We first prove that unital channels cannot decrease the entropy of a state, by considering the relative entropy
\begin{equation}
    D(\rho || \sigma)=\texttt{Tr}(\rho \text{ln} (\rho)- \rho \text{ln} (\sigma)).
\end{equation}
For any quantum channel it can be shown that $D(\rho || \sigma) \geq D(C[\rho] || C[\sigma])$ \cite{NielsenChuang}. For a unital channel $C$, if we set  $\sigma=\identity/d$, we obtain $D(\rho || \identity/d) \geq D(C(\rho) || \identity/d)$. Writing $C[\rho]=\rho'$, we find
\begin{align}
    D(\rho  || \identity/d)&\geq D(\rho'|| \identity/d) \nonumber \\
    \implies \texttt{Tr}(\rho \text{ln} (\rho)- \rho \text{ln} (\identity/d))  &\geq \texttt{Tr}(\rho' \text{ln} (\rho')- \rho' \text{ln} (\identity/d))\nonumber \\
    \implies \texttt{Tr}(\rho \text{ln} (\rho))+ \tr{\rho} \ln d   &\geq \texttt{Tr}(\rho' \text{ln} (\rho')) + \tr{\rho'} \ln d \nonumber \\
    \implies \texttt{Tr}(\rho \text{ln} (\rho)) &\geq \texttt{Tr}(\rho' \text{ln} (\rho')) \nonumber \\
    \implies S(\rho') &\geq S(\rho)
\end{align}
Hence for unital channels $W^{\mathrm{total}}_{C} =0$. To prove the converse, note that a non-unital channel must satisfy $C[\identity/d] \neq \identity/d$. As the maximally mixed state $\identity/d$ is the unique state with maximal entropy, it must therefore be the case that $S(C[\identity/d]) < S(\identity/d)$ and hence  $W^{\mathrm{total}}_{C} >0$.

\subsection{Thermodynamic costs}

Given that many channels provide a work benefit, with heat transformed into work as the system undergoes a cyclic process, it is interesting to explore consistency with the second law of thermodynamics. 

To investigate this, we construct an explicit model of the channel, via a unitary interaction between the target system and an additional ancilla. We will label this additional ancilla (which forms part of the device which implements the channel) by $Z$ to distinguish it from the ancilla $A$ that forms part of the original system. Indeed, any channel can be written as 
\begin{equation} \label{eq::chanimplicit}
C[\rho_t] = \trs{V \rho_t \otimes \rho_{z} V^{\dagger}}{z} 
\end{equation} 
where $\rho_{z}$ is the initial state of the additional ancilla and $V$ is unitary. For simplicity, we will assume that $Z$ is degenerate, with $H_z=0$.

In the same way as for \eqref{eq:entropyincrease}, due to the subadditivity of the entropy we find that 
\begin{align} 
\Delta S_t + \Delta S_z \geq 0. 
\end{align} 
For any given protocol, from \eqref{eq:worktotal} we have $W^{\mathrm{total}}_{C,P} \leq -T \Delta S_t$. 

If we wanted to reset $Z$ to its initial state at the end of the protocol, via a procedure involving the thermodynamic bath,it follows from \eqref{eq:work_extraction} that we would gain work 
\begin{align} 
W^{\mathrm{reset}}_{C,P} \leq - T \Delta S_z,
\end{align} 
where $\Delta S_z$ refers to the entropy change of $Z$ when the channel was originally implemented. Hence the total work gained by performing the protocol and then resetting the device satisfies
\begin{align} \label{eq:hiddencosts}
W^{\mathrm{total}}_{C,P} + W^{\mathrm{reset}}_{C,P}  \leq 0
\end{align} 
in accordance with the second law. 

Note that we have not considered the costs associated with generating the initial state of the system or device ($\rho_s$ and $\rho_z$) before the protocol begins. In the former case, this is because the system is returned in the same state at the end. In the latter case, this is part of the resource that is the device, and we consider in this section the cost of resetting it. However, there is an interesting subtlety that arise if either of these states is not full rank, because then it may be impossible to reconstruct them exactly via the thermodynamic operations in our framework. For the system, by considering a series of full-rank initial states that comes arbitrarily close to the optimal state, we can see that the supremum in theorem \ref{thrm::workbenefitC} indeed becomes the maximum over all states. We can also consider a series of devices, for each of which $\rho_z$ is full rank, which implement channels as close as desired to any desired channel.

\subsection{Catalytic channels}

A particularly interesting class of channels in this context are catalytic channels, which are channels that can be implemented in such a way that $Z$ remains unchanged.  
\begin{definition} A channel $C$ is catalytic if there exists a unitary $V$ and state $\rho_z$ such that 
\begin{align} 
C[\rho_t] &= \trs{V \rho_t \otimes \rho_z V^{\dagger}}{z} \\
\rho_z &= \trs{V \rho_t \otimes \rho_z V^{\dagger}}{t}
\end{align}
\end{definition} 
Given that a catalytic implementation of a channel  does not need resetting afterwards, it must satisfy $W^{\mathrm{reset}}_{C,P} =0$. Hence from \eqref{eq:hiddencosts} catalytic channels cannot provide a work benefit. From Corollary \ref{cor:unital} we therefore discover that catalytic channels must be unital. 
This raises the question of whether all unital channels are catalytic. In appendix \ref{app::unital} we show that this is not the case. There are some channels which are unital, and thus useless for work extraction, but nevertheless lack a catalytic implementation. A particular example is the Werner-Holevo channel for a three-dimensional target system:
\begin{equation}\label{eq:WHchan}
    \rho \rightarrow C^{\text{WH}}(\rho)=\frac{1}{2}(\tr{\rho}\identity-\rho^T), 
\end{equation}
where $\rho^T$ corresponds to the transpose of $\rho$ in a particular basis. We have therefore proven 
\begin{corollary} All catalytic channels are unital. However, there exist unital channels which are not catalytic.  
\end{corollary} 
Interestingly, while we have used thermodynamic arguments to derive this result, it relates two properties of channels which are not intrinsically thermodynamic. 

Overall, we have shown the following triad of implications between quantum channels that are not useful for generating work, channels that have a catalytic implementation, and the set of unital channels;
\begin{equation}
    \text{Catalytic} \implies \text{No work benefit} \iff \text{Unital}
\end{equation}

\subsection{Channels using explicit batteries}

To provide further justification for the implicit battery results obtained earlier, we now show how they can be extended to include explicit batteries. 

As in \eqref{eq::chanimplicit}, the channel is implemented by acting with a unitary transformation $V$ on the target system and an ancilla initially in state $\rho_z$. The ancilla is internal to the device implementing the channel, and for simplicity we assume it is degenerate in energy. 

We can expand $V$ in an energy basis as 
\begin{equation}\label{eq::unitary}
   V=\sum_{a b c d}\alpha_{a b}^{c d}\ket{a}\bra{b} \otimes \ket{c} \bra {d}
\end{equation}
where $\alpha_{a b}^{c d}$ are complex numbers describing state transition amplitudes within the Hilbert spaces of the system and ancilla.

The unitary $V$ may not in general be energy conserving, and as such the resultant channel may either raise or lower the internal energy of the target system. To move to the \textit{explicit} battery case, we construct an energy conserving unitary $\tilde{V}$, which  draws any such energy change from the battery. In particular, whenever the energy of the system changes, we shift the position weight by the same amount in the opposite direction. More concretely 
\begin{definition}
The \textit{extension} of a unitary used to implement a quantum channel in the implicit battery case (equation \eqref{eq::unitary}) is given by
\begin{equation}\label{eq::implement}
    {\tilde{V}}=\sum_{a b c d}\alpha_{a b}^{c d}\ket{a}\bra{b} \otimes \ket{c} \bra {d} \otimes \Gamma_{E_b-E_a}.
\end{equation}
where $\Gamma$ is the translation operator for the  weight, defined in \eqref{eq::translation}. 
\end{definition}

It is straightforward to check that $\tilde{V}$ is unitary, commutes with the total Hamiltonian of the system, ancilla and weight, and has the weight translation invariance property that we assume in the explicit battery framework

As the device interacts with the external weight, it's action on the target system will in general depend on the state of the weight. In particular, the channel implemented given a weight state $\rho_w$ is 
\begin{equation}
    C_{\rho_w}[\rho_t]=\texttt{Tr}_{z w}[{\tilde{V}}(\rho_t \otimes \rho_z \otimes \rho_w){\tilde{V}}^{\dagger}].
\end{equation}
However, as long as the weight is prepared in a state with a narrow spread in momentum about $p=0$, the channel implemented using the explicit battery will be very close to the desired channel (see appendix \ref{app::secondlaw}), and can be made arbitrarily close by choosing an appropriate initial weight state. Note that such initial states of the weight are also those required to perform arbitrary transformations of a system and bath with optimal thermodynamic efficiency. 

As a simple concrete example, consider the initial weight state 
\begin{equation} \label{eq::tophat} \ket{\psi_L}=\frac{1}{\sqrt{2L}}\int_{-L}^Ldx \ket{x} \end{equation} with $\rho_w=\ket{\psi_L}\bra{\psi_L}$. Then it follows from the results of appendix \ref{app::secondlaw} that 
\begin{equation} 
\lim_{L\rightarrow \infty} C_{\rho_w}[\rho_t] = C[\rho_t]
\end{equation} 
For a weight state with very large $L$, the effect of the channel on the system will be very close to that in the implicit battery case. This means that it's change in average energy will also be very similar to $\Delta U_{\rho_t \rightarrow C[\rho_t]}$ 
As energy is conserved overall the shift in the average energy of the weight will therefore be very close to that given by definition that given by Definition \ref{def::implicitworkcostC} as desired. 

\section{Measurements}\label{sec::measurements}

We now consider the thermodynamic benefit of a general quantum measurement. This can be described by assigning to each outcome $i$ of the measurement a completely-positive transformation $C_i$. The probability of obtaining a particular outcome when the measurement is performed on a state $\rho$ is 
\begin{equation} 
p_i = \tr{C_i[\rho]}
\end{equation} 
with the state after measurement being given by 
\begin{align} 
\rho_i = \frac{C_i[\rho]}{p_i}
\end{align} 
Together, summing over all outcomes of the measurement must yield a trace-preserving channel 
\begin{equation}
    C[\rho]=\sum_i C_i[\rho].
\end{equation}
For convenience, we will also define $C_{s,i}$, corresponding to applying $C_i$ to the target system, and the identity channel to the ancillary system.

As before, we consider that any energy required for the device to implement the measurement is provided by the same battery as is used to perform  thermodynamic operations within the framework. 

A general protocol $P$ for work extraction in this case is similar to before, but the resetting operation will in general depend on the measurement result.  
\begin{enumerate} 
\item The system and bath are initialized in an uncorrelated initial state  $\rho=\rho_s \otimes \tau_b$. 
\item The measurement is applied to the target system, giving result $i$ with probability $p_i$, and leaving the state  as $\sigma_{i} =  (C_{s,i}[\rho_s] \otimes \tau_b)/p_i $.
\item For each result $i$, a unitary interaction $U_i$ is performed, transforming the state into $\sigma_{i}' = U_i \sigma_i U_i^{\dagger}$, where $\trs{\sigma'_{i}}{b} = \rho_s$ (i.e. the system returns to its initial state for each result).
\end{enumerate} 

Specifying a protocol $P$ includes choosing the desired ancillary and bath systems to complement the target system, as well as the initial state $\rho_s$ of the system and the set of unitaries $U_i$. 

As before, the total work extracted by the protocol $P$ is given by the sum of the work gained in each step. As there is no interaction with the bath, the average work gain when applying the measurement must be equal to the average energy loss of the system.  
\begin{definition} \label{def::implicitworkcostM}
The work gained on average when applying the measurement $\{C_i\}$ to the target system in a state $\rho_t$ (part of a larger system in state $\rho_s$)  is given by
\begin{align} 
W^{\mathrm{apply}}_{\{C_i\}} &\equiv - \sum_i p_i \, \Delta U_{\rho_t \rightarrow \rho_{t,i}} \nonumber \\ 
&=  \sum_i p_i\, \tr{ H_t \rho_t - H_t   \frac{C_i[\rho_t]}{p_i\,} } \nonumber \\
& = \tr{ H_t \rho_t - H_t C(\rho_t) }\nonumber \\
& = W^{\mathrm{apply}}_C
\end{align} 
where $C[\rho] = \sum_i C_i[\rho]$.
\end{definition}

As before the total work benefit provided by the measurement is obtained by summing the average work benefit in each step
\begin{align} 
W^{\mathrm{total}}_{\{C_i\},P} = W_1 + W_2 + W_3 
\end{align} 
Using the above definition and earlier results, the work gain in each step can be bounded as follows:
\begin{align} 
W_1 &=0 \\
W_2 &= W^{\mathrm{apply}}_{\{C_i\}} =  E(\rho_s) - \sum_i p_i\, E(\sigma_i) \\ 
W_3 &\leq \sum_i p_i \left( F(\sigma_i) - F(\rho_s) \right). 
\end{align} 
Hence using $F(\rho) = E(\rho) - T S(\rho)$,
\begin{align} \label{eq:worktotalM} 
W^{\mathrm{total}}_{\{C_i\},P} &\leq T \sum_i p_i \left( S(\rho_s) - S(\sigma_{s,i}) \right) \nonumber  \\
&\leq T \sum_i p_i \left( S(\rho_t) - S(\sigma_{t,i}) \right) \nonumber \\
 &= T \left( S(\rho_t) - \sum_i p_i S(\sigma_{t,i}) \right) 
\end{align} 
where $\sigma_{s,i}= \trs{\sigma_i}{b} = \frac{C_{s,i}[\rho_s]}{p_i} $, and $\sigma_{t,i}= \trs{\sigma_{s,i}}{a} = \frac{C_{i}[\rho_t]}{p_i} $. As in the case of channels, the last step follows from the monotonicity of the mutual information under the action of local channels. However, this case is more complicated and is shown in appendix \ref{app::workevaluations}.

Note  that for any $\rho_t$ we can consider a protocol in which $\rho_s= \rho_t$ (i.e. no ancillas are used), and each $U_i$ is a thermodynamically efficient protocol for transforming $\sigma_{t,i}$ into $\rho_t$. Hence there exist protocols which come as close as desired to saturating \eqref{eq:worktotalM} for any $\rho_t$. 

Finally, we optimize over protocols to find the work benefit of the measurement $\{C_i\}$.
\begin{theorem}\label{thrm::workbenefitM}
The  work benefit of the measurement $\{C_i\}$ is given by 
\begin{align}\label{thrm::measurments}
 W^{\mathrm{total}}_{\{C_i\}} &\equiv \sup_P\, W^{\mathrm{total}}_{\{C_i\},P} = \max_{\rho_t} T \left( S(\rho_t) - \sum_i p_i S(\sigma_{t,i}) \right)
\end{align}
\end{theorem} 

Note that this means that any complete basis measurement has a work benefit equal to $T \ln d_s$, taking $\rho_t$ as the maximally mixed state in the maximization above, and noting that each post-measurement state is pure and thus has entropy zero. 

Furthermore, we can now make the statement that `measurements are of more thermodynamic use than channels'. Of course, to make this statement precise we associate a channel $C$ to a collection of measurement sub channels, $C[\rho]=\sum_i C_i[\rho]$. Then from theorem \ref{thrm::workbenefitC}, theorem \ref{thrm::workbenefitM} and the concavity of the entropy we find that $ W^{\mathrm{total}}_{C} \leq  W^{\mathrm{total}}_{\{C_i\}}$, so measurements are capable of providing more work than a channel on the average. The way to see this is that we may construct an adaptive resetting protocol for the measurement case, which allows us an extra level of control over the state based on informative measurement outcomes.

\subsection{Thermodynamic costs} 

When considering this work benefit in the context of the second law of thermodynamics, we again find that the cost of resetting the measurement device and erasing all information about the measured outcome would carry a thermodynamic cost which outweighs the work benefit. This was previously observed in the solution to Maxwell's demon \cite{Lloyd1997}. 

In particular, we can represent the measurement by a unitary interaction $V$ between the target system and an additional ancilla inside the measurement device denoted by $Z$, in initial state $\rho_z$. Each measurement result $i$ corresponds to an orthogonal projector $\Pi_i$ on $Z$, such that 
\begin{equation}
p_i = \tr{V (\rho_t \otimes \rho_z) V^{\dagger} (I \otimes \Pi_i) }
\end{equation} 
and 
\begin{equation} 
\sigma_{t,i} = \frac{1}{p_i} \trs{V (\rho_t \otimes \rho_z) V^{\dagger} (I \otimes \Pi_i) }{z}.
\end{equation} 
we can think of the conditional unitary applied in step 3 of the protocol, which depends on the measurement result as 
\begin{equation} 
U = \sum_i U_i \otimes \Pi_i
\end{equation}
If the measurement device $Z$ is reset to its initial state at the end of the protocol, via a unitary interaction with the bath, then overall we have constructed a fully unitary protocol on $T$, $B$ and $Z$ which returns both $T$ and $Z$ to their initial states. We have previously proven that this can only lead to a net decrease of work. For more on the resource costs of preparing the initial state of the measuring device and connections with the third law of thermodynamics, see \cite{Guryanova2020}.

Note that if one includes the experimenter as an additional agent inside the protocol, who reads the measuring device and implements step 3 and resets the state of the device based on their knowledge of the measurement result, then this process may appear to violate the second law. However, in this case, one should also reset the memory of the experimenter at the end of the protocol. This would carry an additional thermodynamic cost which restores compatibility with the second law. 

\subsection{Conditional work benefit} 
For a single channel which transforms $\rho_t$ into $\sigma_t$, we can obtain an optimal work benefit arbitrarily close to $T(S(\rho_s) - S(\sigma_t))$. For a measurement the corresponding optimal work benefit is $\sum_i p_i T (S(\rho_s) - S(\sigma_{t,i})$. Given these results, it would be natural to assume that the optimal conditional work benefit for a measurement given that result $i$ was obtained is $ T(S(\rho_s) - S(\sigma_{t,i}))$.

However, if we consider how the measurement would be implemented with an explicit battery, we find that this is not the case. To illustrate this, let us consider the simple example of an energy measurement on a qubit. Imagine that the target system has Hamiltonian 
\begin{equation}
    H_t=E\ket{1}\bra{1}.
\end{equation}
and is initially in a maximally mixed  state
\begin{equation}
\rho_t = \frac{1}{2} (\proj{0} + \proj{1}),  
\end{equation} 
and that the measurement is characterised by 
\begin{align} 
C_0[\rho] = \proj{0} \rho \proj{0} \nonumber \\
C_1[\rho] = \proj{1} \rho \proj{1}.
\end{align} 
We can perform the measurement by preparing a `measurement device' qubit (with zero Hamiltonian) in the state $\rho_z = \proj{0}$ and implementing a `controlled-NOT' unitary transformation from the target  onto the device.
\begin{align}
V = \proj{0} \otimes I + \proj{1} \otimes X 
\end{align} 
where $X = \op{0}{1} + \op{1}{0}$. Note that in this case $V$ commutes with the total Hamiltonian, and therefore requires no interaction with the battery to perform. Hence $W^{\textrm{apply}}_{\{C_0, C_1\}}=0$. 

In the case that the measurement result is 0, we use the thermodynamic bath to  unitarily transform the system's final state $\sigma_{t,0} = \proj{0}$  back into its initial state $\rho_t$, obtaining an optimal amount work given by the reduction in free energy. In this particular case  $F(\sigma_{t,0})= 0$ and $F(\rho_t) = E/2 - T \ln 2$, hence the optimal conditional work benefit given result 0 is
\begin{align} 
W^{\textrm{total}}_{\{C_0, C_1\} | 0 } =  T \ln 2 - E/2. 
\end{align} 
If result 1 is obtained then $\sigma_{t,1} = \proj{1}$ with $F(\sigma_{t,1})= E$. Hence the conditional work benefit for this result is
\begin{align} 
W^{\textrm{total}}_{\{C_0, C_1\} | 1 } = T \ln 2 + E/2. 
\end{align} 
As each case occurs with probability 1/2, overall the total work benefit is 
\begin{align} 
W^{\textrm{total}}_{\{C_0, C_1\}} = T \ln 2.
\end{align} 
Note that although this is equal to 
$\sum_i p_i T (S(\rho_s) - S(\sigma_{t,i})$
the conditional work benefit for result $i$ is not equal to $S(\rho_s) - S(\sigma_{t,i}) = T \ln 2$. 

The situation becomes considerably more complicated when the unitary $V$ implementing the measurement does not commute with the Hamiltonian of the system. In such cases, in order to account for the energy change of applying the measurement, we extend the unitary in a similar way to \eqref{eq::implement} to incorporate the explicit battery.

The conditional work benefit of applying a measurement  given the broad `top-hat' wavefunction for the weight considered in \eqref{eq::tophat} is given by the following theorem
\begin{theorem} \label{thm3}
The conditional work benefit $ W^{\mathrm{apply}}_{\{C_i\}|i}$ of applying a measurement ${\{C_i\}}$ and obtaining result $i$, given that the initial state of the weight is $\ket{\psi_L}=\frac{1}{\sqrt{2L}}\int_{-L}^Ldx \ket{x}$, and omitting terms of $O\left(\frac{1}{L}\right)$ (i.e. taking the large $L$ limit) is 
\begin{equation}\label{eq::thrm3}
     W^{\mathrm{apply}}_{\{C_i\}|i}=\frac{1}{p_i} \tr{{C}_i \left[ \frac{H_t\rho_t+\rho_t H_t}{2}\right] - H_t {C}_i[\rho_t]} 
\end{equation}
\end{theorem} 
which we prove in appendix \ref{app::workevaluations}. Note that if we average over the conditional work benefit for each measurement outcome, we recover $W^{\mathrm{apply}}_{\{C_i\}}$, because 
\begin{align} 
 \sum_i p_i  W^{\mathrm{apply}}_{\{C_i\}|i} & = \tr{{C} \left[ \frac{H_t\rho_t+\rho_t H_t}{2}\right] - H_t {C}[\rho_t]}   \nonumber \\
&= \tr{H_t\rho_t - H_t {C}[\rho_t]} \nonumber \\
&=  W^{\mathrm{apply}}_{\{C_i\}}, 
\end{align} 
where the second line follows because $C$ is trace-preserving. 

Including the work benefit of resetting the system to its initial state  we can then calculate the total conditional work benefit of the measurement, given that result $i$ was obtained and the system was prepared in state $\rho_s$. 
\begin{align} 
W^{\mathrm{total}}_{\{C_i\}|i, \rho_s} &= W^{\mathrm{apply}}_{\{C_i\}|i} + F(\sigma_{s,i}) - F(\rho_s) \nonumber \\
&= \frac{1}{p_i} \tr{{C}_i \left[ \frac{H_t\rho_t+\rho_t H_t}{2}\right]} - \tr{H_t \sigma_{t,i}} \nonumber \\ &\qquad + F(\sigma_{s,i}) - F(\rho_s) \nonumber \\
&= \frac{1}{p_i} \tr{{C}_i \left[ \frac{H_t\rho_t+\rho_t H_t}{2}\right]} - \tr{H_t \rho_t} \nonumber \\ &
\qquad + \tr{H_a \sigma_{a,i}} - \tr{H_a \rho_a}  \nonumber \\ &\qquad  +T (S(\rho_s) - S(\sigma_{s,i}))  \label{eq:totalconditional}
\end{align} 
Note that this result does not include small corrections due to the finite width of the weight state, and small inefficiencies in the thermodynamic reset operation, both of which can be made as small as desired.

The same results as \eqref{eq::thrm3} and \eqref{eq:totalconditional} can be be obtained for other initial states of the weight, including a Gaussian of standard deviation $L$. However, in general they can depend on the initial state of the weight. In particular, consider  an asymmetric `triangular' initial state of the weight 
\begin{equation}
    \ket{\tau_L} = \sqrt{\frac{3}{L^3}}\int_{-\frac{3L}{4}}^{\frac{L}{4}}(x+\frac{3L}{4})\ket{x}dx,
\end{equation}
where the constants are chosen such that this is normalised and has average energy zero.
We show in appendix \ref{app::otherweights} that this initial state of the weight leads to a different result for $ W^{\mathrm{apply}}_{\{C_i\}|i}$, even in the limit as $L \rightarrow \infty$. This is different to our earlier results, which only required that the initial weight state had an initial momentum close to zero with high probability.

\section{Post-selected Measurements}\label{sec::post}

Post-selection is the process of discarding statistics in a data set, and has been shown to elude to many interesting foundational questions \cite{Aharonov1964} and is of practical importance in weak measurement \cite{Aharonov1988}. In this section, we will consider the work benefit of making post-selected measurements. 

Given a measurement represented by a set of completely positive maps $\{C_i\}$, we denote some subset of the measurement outcomes $(i \in \text{succ})$ as indicating success in the post selection, while the others  $(i \in \text{fail})$ correspond to failure. When considering the work benefit of the post-selected measurement, we  only consider cases in which the post-selection succeeds. 

It is convenient to define the probability of success in the post-selection as  $p_{\text{succ}} = \sum_{i \in \text{succ}} p_i$, where $p_i = \tr{C_i [\rho]}$ as before. The total work benefit of the post-selected measurement can then be calculated using the conditional work benefits in the previous section. In particular,
\begin{equation} 
W^{\mathrm{total}}_{\{C_i\}|\text{succ}, \rho_s} = \frac{1}{p_{\text{succ}}} \sum_{i \in \text{succ}}  p_i W^{\mathrm{total}}_{\{C_i\}|i, \rho_s}
\end{equation} 
From \eqref{eq:totalconditional}, for a broad top-hat initial state of the weight and efficient thermodynamic transformations, we  obtain 
\begin{align} \label{eq:generalpost} 
W^{\mathrm{total}}_{\{C_i\}|\text{succ}, \rho_s} &=  \sum_{i \in \text{succ}} \tr{{C}_i \left[ \frac{H_t\rho_t+\rho_t H_t}{2\, p_{\text{succ}} }\right]} - \tr{H_t \rho_t}
 \nonumber \\ &
\qquad + \sum_{i \in \text{succ}} \frac{p_i}{p_{\text{succ}}} \tr{H_a \sigma_{a,i}} - \tr{H_a \rho_a}  \nonumber \\ &\qquad  +T \left(S(\rho_s) - \sum_{i \in \text{succ}} \frac{p_i}{p_{\text{succ}}}  S(\sigma_{s,i})\right).
\end{align}

In the simple case in which the measurement has only two outcomes, one corresponding to success and the other to failure, given by completely positive maps $\{ C_{\text{succ}}, C_{\text{fail}} \}$ we find  
\begin{align} \label{eq:binarypost} 
W^{\mathrm{total}}_{\{ C_{\text{succ}}, C_{\text{fail}} \}|\text{succ}, \rho_s} &=   \tr{{C}_{\text{succ}} \left[ \frac{H_t\rho_t+\rho_t H_t}{2\, p_{\text{succ}} }\right]} \nonumber \\ & \qquad - \tr{H_t \rho_t}
 \nonumber \\ &
\qquad +  \tr{H_a \sigma_{a,\text{succ}}} - \tr{H_a \rho_a}  \nonumber \\ &\qquad  +T \left(S(\rho_s) -  S(\sigma_{s,\text{succ}})\right).
\end{align}

Given any post-selected measurement $\{C_i\}$, we can always consider the corresponding binary outcome measurement in which we only record whether the post-selection succeeds or fails. This would be given by $\{C_{\text{succ}},C_{\text{fail}} \}$, where $C_{\text{succ}} = \sum_{i \in \text{succ}} C_i$ and $C_{\text{fail}} = \sum_{i \in \text{fail}} C_i$. In this case, one can see by comparing \eqref{eq:generalpost}  and \eqref{eq:binarypost} and using the concavity of the entropy that $W^{\mathrm{total}}_{\{C_i\}|\text{succ}, \rho_s} \geq W^{\mathrm{total}}_{\{ C_{\text{succ}}, C_{\text{fail}} \}|\text{succ}, \rho_s}$ as expected.

\subsection{Every post-selected channel can lead to unbounded amounts of work} \label{sec::postselection} 

We will now show that any post-selected measurement where the post-selection is non-trivial leads to an unbounded work benefit.  Hence even one copy of a post-selected measurement can be used to generate arbitrarily large amounts of work in our thermodynamic framework. 

\begin{theorem} 
Any non-trivial post-selected measurement  (i.e. where the success probability has some dependence on the state) can be used to obtain an unbounded work benefit. 
\begin{align} 
 W^{\mathrm{total}}_{\{C_i\}|\text{succ}} = \sup_{\rho_s} W^{\mathrm{total}}_{\{C_i\}|\text{succ}, \rho_s} = \infty. 
 \end{align}
 \end{theorem} 

To prove this, we first note that as the work benefit for any multi-outcome measurement $\{ C_i \}$ is more than the work benefit for the corresponding binary outcome measurement, $\{C_{\text{succ}},C_{\text{fail}} \}$ in which we only consider success or failure, it suffices to consider binary measurements.

To illustrate the key ideas of the proof, let us first consider a simple example in which 
\begin{align} 
C_{\text{succ}}[\rho] = \proj{0} \rho \proj{0} \nonumber \\
C_{\text{fail}}[\rho] = \proj{1} \rho \proj{1}.
\end{align} 
and the system has zero Hamiltonian $H_s=0$. We then apply this post-selected measurement to the state 
\begin{equation}
    \rho_s=\frac{1}{2} \proj{0}_t \otimes \ket{\phi}\bra{\phi}_a + \frac{1}{2} \proj{1}_t \otimes \frac{1}{d_a}\mathbf{1}_a
\end{equation}
where $\ket{\phi}$ is any pure state of the ancilla, and $d_a$ is the ancilla dimension. If the post-selection is successful, the post-measurement state is $\sigma_{s, \text{succ}} = \proj{0} \otimes \proj{\phi}$. This is a zero entropy state, which can be expanded using unitary operations on the system and a bath back to the initial state $\rho_s$. This allows us to extract a total amount of work arbitrarily close to  
\begin{align} 
W &= F(\sigma_{s, \text{succ}}) - F(\rho_s) \nonumber \\
& = T(S(\rho_s) - S(\sigma_{s, \text{succ}})) \nonumber \\
&=  T \left(\ln 2 + \frac{1}{2}\ln d_a \right).
\end{align} 
Since  there  is  no  restriction on the dimension of the ancilla $d_a$, and this work grows as $\ln d_a$, we can extract an unbounded amount of work by considering larger and larger ancillas. 

Now let us generalise this approach to an arbitrary post selected measurement $\{C_{\text{succ}},C_{\text{fail}} \}$. Given any completely positive map $C_{\text{succ}}$, there exists a positive operator $M_{\text{succ}}$ such that $p_{succ}  = \tr{C_{\text{succ}}[\rho_t]} = \tr{M_{\text{succ}} \rho_t}$ \footnote{Writing $C_{\text{succ}}$ in Kraus decomposition as $C_{\text{succ}}[\rho] = \sum_j K_j \rho K_j^{\dagger}$, we have  $M_{\text{succ}} = \sum_j K_j^{\dagger} K_j$}. Let  $\ket{u}$ and  $\ket{v}$ be eigenvectors of  $M_{\text{succ}}$ with the maximal and minimal eigenvalue respectively,
\begin{align} 
M_{\text{succ}}\ket{u}& =\lambda_{\text{max}}\ket{u},\\ M_{\text{succ}}\ket{v}& =\lambda_{\text{min}}\ket{v}.
\end{align} 
Note that in order for the post-selection to be non-trivial, we require that $\lambda_{\text{max}} > \lambda_{\text{min}}$. If this is not the case then $M_{\text{succ}}$ is proportional to the identity and all states would give the same probability of success. For any such measurement, $C_{\text{succ}} = \alpha C$ where $\alpha$ is a constant and $C$ is a trace-preserving channel. This corresponds to performing $C$ and then  failing independently at random with probability $(1-\alpha)$, which for all practical purposes is the same as just performing $C$

Now consider applying the measurement to the state 
\begin{equation} \label{eq::post-selectedstate} 
\rho_s = \frac{1}{2} \proj{u}_t \otimes \proj{\phi}_a  + \frac{1}{2} \proj{v}_t\otimes \frac{1}{d_a}\mathbf{1}_a,
\end{equation}
where we take the ancilla to have zero Hamiltonian ($H_a=0)$.
From \eqref{eq:binarypost}, the total work benefit in this case is given by
\begin{align} 
W^{\mathrm{total}}_{\{ C_{\text{succ}}, C_{\text{fail}} \}|\text{succ}, \rho_s} &=   \tr{L_{\text{succ}} \left( \frac{H_t\rho_t+\rho_t H_t}{2\, p_{\text{succ}} }\right)} \nonumber \\ & \quad - \tr{H_t \rho_t}
 \nonumber \\ &
\quad +  \tr{H_a \sigma_{a,\text{succ}}} - \tr{H_a \rho_a}  \nonumber \\ &\quad  +T \left(S(\rho_s) -  S(\sigma_{s,\text{succ}})\right)
\end{align} 
We can use the cyclic symmetry of the trace and the definition of $\ket{u}$ and $\ket{v}$, as well as the fact that $p_{\text{succ}} = (\lambda_{\text{max}} + \lambda_{\text{min}})/2$ to simplify the first term on the right-hand side. Also using $H_a =0$ and $S(\rho_s) = \ln 2 + \frac{1}{2} \ln d_a$  gives 
\begin{align} 
W^{\mathrm{total}}_{\{ C_{\text{succ}}, C_{\text{fail}} \}|\text{succ}, \rho_s}& = \frac{\lambda_{\text{max}} \bra{u} H_t \ket{u} + \lambda_{\text{min}} \bra{v} H_t \ket{v}}{\lambda_{\text{max}} + \lambda_{\text{min}}} \nonumber \\
& \quad - \frac{ \bra{u} H_t \ket{u} + \bra{v} H_t \ket{v}}{2} \nonumber \\
& \quad + T \left(\ln 2 + \frac{1}{2} \ln d_a -  S(\sigma_{s,\text{succ}})\right)
\end{align}
Note that $ \frac{\lambda_{\text{max}} \bra{u} H_t \ket{u} + \lambda_{\text{min}} \bra{v} H_t \ket{v}}{\lambda_{\text{max}} + \lambda_{\text{min}}}$ and $\frac{ \bra{u} H_t \ket{u} + \bra{v} H_t \ket{v}}{2}$ both represent convex mixtures of expected energies. They must therefore lie within the largest and smallest eigenvalues of $H_t$, which we denote respectively by $E_{\text{max}}$ and $E_{\text{min}}$. Hence 
\begin{align} 
W^{\mathrm{total}}_{\{ C_{\text{succ}}, C_{\text{fail}} \}|\text{succ}, \rho_s}& \geq E_{\text{min}} - E_{\text{max}} \nonumber \\ &\quad + T \left(\ln 2 + \frac{1}{2} \ln d_a -  S(\sigma_{s,\text{succ}})\right)
\end{align}

Furthermore,  from the subadditivity of the entropy we have 
\begin{equation}
S(\sigma_{s,\text{succ}}) \leq S(\sigma_{t,\text{succ}})+S(\sigma_{a,\text{succ}})\leq \ln(d_t) + S(\sigma_{a,\text{succ}})
\end{equation}
Defining 
\begin{equation} 
q= \frac{\lambda_{\text{min}}}{\lambda_{\text{max}} + \lambda_{\text{min}}},
\end{equation} 
note that
\begin{align}
\sigma_{a,\text{succ}} = (1-q) \proj{\phi} + q\left( \frac{1}{d_a}\mathbf{1}_a \right)
\end{align} 
Using the fact that $S(\sum_i p_i \rho_i) \leq H(p_i) + \sum_i p_i S(\rho_i)$, where $H(p_i)$ is the Shannon entropy, we have 
\begin{align} 
S(\sigma_{a,\text{succ}}) \leq H(q) + q \ln d_a \leq \ln 2 + q \ln d_a
\end{align} 
Putting everything together we then have 
\begin{align} 
W^{\mathrm{total}}_{\{ C_{\text{succ}}, C_{\text{fail}} \}|\text{succ}, \rho_s}& \geq E_{\text{min}} - E_{\text{max}} \nonumber \\ &\quad + T \left(\left(\frac{1}{2}-q \right) \ln d_a - \ln d_t \right).
\end{align}
As $\lambda_{\text{max}} > \lambda_{\text{min}}$, we have $q < \frac{1}{2}$. Hence the work benefit  grows in an unbounded way with $\ln d_a$. 

This result was derived for a top-hat weight state and does not include $O(1/L)$ corrections due to the finite width of that state, and $O(\epsilon)$ corrections due slight inefficiencies in the thermodynamic protocol for resetting the system state. However, both of these can be made as small as desired. Using other weight states (assuming they are sharply peaked around momentum zero) would also not change the scaling with $d_a$. 

\section{Conclusions}\label{sec::conclusions}

We have analysed the work benefit of a single use of a general channel or measurement within a framework for quantum thermodynamics in which work corresponds to a change in average energy of an \textit{implicit} or \textit{explicit} battery. 

For the case of channels, we find that the  work benefit depends on the maximum reduction in entropy that the channel can induce on the target system. Obtaining this maximal work generally requires a large thermodynamic bath to efficiently return the system to its initial state, and in the explicit battery model an initial state of the weight which is narrowly peaked about zero in momentum space. 
A consequence of this is that a channel is useful for work extraction if and only if it is unital. 
We also show some interesting relationships between unit channels, catalytic channels and work extraction.

We have also considered the case of quantum measurements. It has been shown that a quantum measurement is a more thermodynamically useful object than the equivalent measure-and-forget type channel, due to the existence of an adaptive resetting protocol in the cycle. We obtain a formula for the conditional work extraction, which is the amount  the battery system changes in average energy, given that we make a measurement and observe a particular outcome. Surprisingly, the conditional work extraction formula has a stronger dependence on the initial state of the battery system than our earlier results, which merits further investigation. This might also have implications for pointer states in the weak measurement formalism \cite{Aharonov1990}.  

We  then investigated the thermodynamics of post-selected measurements, which have been shown to be extra-ordinarily powerful. Given a single use of any measurement with non-trivial post-selection,  it is possible to extract an unbounded amount of work  within a single thermodynamic cycle. This might herald interesting avenues of investigation if one takes the view that there exists fundamental post-selection in nature \cite{Aharonov2017}, such as in black-hole physics \cite{Horowitz2004,Harlow2016}, and analyses the thermodynamics of such theories.

\acknowledgements{The authors acknowledge helpful conversations with S. Popescu. TP acknowledges support from the EPSRC.}

\bibliographystyle{unsrt}
\bibliography{main}

\appendix

\section{The second law for explicit batteries} \label{app::secondlaw} 

In order to prove that the second law of thermodynamics, and equation \eqref{eq:work_extraction}, hold in the case of explicit batteries, we first note that as allowed unitaries commute with the translation operator on the weight, they can be written in the form
\begin{equation} 
\tilde{U}  = \int dp\; U(p) \otimes \proj{p}_w
\end{equation}
where each $U(p)$ is a unitary on the system and bath. This means that the system and bath transform via a mixture of unitaries.
\begin{align} 
\rho'_{sb} &= \trs{ \tilde{U} (\rho_{sb} \otimes \rho_w)  \tilde{U}^{\dagger} }{w} \nonumber \\ 
&= \int dp\;  \mu(p) U(p) \rho_{sb} U(p)^{\dagger}, \label{eq:mixunitaries}  
\end{align}
where $\mu(p) = \tr{\rho_w \proj{p}} $ is the probability density of weight momentum. 
Due to the concavity of the entropy, a mixture of unitaries can only increase the entropy of the system and bath, hence $S(\rho'_{sb}) \geq S(\rho_{sb})$. Combining this with the subadditivity of the entropy and the fact that the initial state is a product state, we obtain 
\begin{equation}
    S(\rho_s)+S(\tau_b) = S(\rho_{sb}) \leq S(\rho'_{sb}) \leq S(\rho_s')+S(\rho_b')
\end{equation}
and the proof of the second law and $W\leq - \Delta F_s$ follows as from \eqref{eq::entropy}. 

To prove that any transformation satisfying this relation can be approximately implemented in the explicit battery case, we first identify a unitary $U$ which approximately achieves the desired transformation in the implicit battery case. $U$ can be expanded as 
\begin{equation} 
U = \sum_{i,j} U_{ij} \op{i}{j},
\end{equation} 
where $\ket{i}$ form an energy eigenbasis for the system and bath with corresponding energies $E_i$. 

The corresponding transformation  in the explicit battery case is given by the unitary 
\begin{equation} 
\tilde{U} = \sum_{i,j} U_{ij} \op{i}{j} \otimes \Gamma_{E_j - E_i}
\end{equation} 
on the system bath and weight. This will give very similar results to the implicit battery case whenever the initial state of the weight has a narrow spread in momentum around zero momentum\footnote{As an example of such a state, consider a broad coherent state of energy, $\ket{\psi_L} = \sqrt{\frac{1}{2L}} \int_{-L}^{L} dx \ket{x}$ for large $L$. Note that such a state satisfies $\Gamma_{E_j - E_i}\ket{\psi_L} \simeq \ket{\psi_L}$. }, or if the system and bath are diagonal in an energy basis and the transformation $U$ is a permutation of those energy states. 

To show this in more detail, we first write $\tilde{U}$ in the momentum basis as 
\begin{equation} 
\tilde{U} = \int dp \; \sum_{i,j} U_{ij} e^{-i (E_j-E_i) p} \op{i}{j} \otimes \proj{p}.
\end{equation} 
From \eqref{eq:mixunitaries}, the final state of the system and bath after applying $\tilde{U}$ is given by 
\begin{equation} \label{eq:finalstateexplicit} 
\rho_{sb}' = \int dp\;  \mu(p) U(p) \rho_{sb} U(p)^{\dagger}.
\end{equation}
where
\begin{equation} \label{eq:explicitU(p)}
U(p) =  \sum_{i,j} U_{ij} e^{-i (E_j-E_i) p} \op{i}{j}. 
\end{equation}
Note that if the initial state of the system and bath is diagonal in energy (i.e. $\rho_{sb} = \sum_n p_n \proj{n}$) and the unitary $U$ is a permutation of those energy levels (i.e. $U_{ij} = \delta_{i, \Pi[j]}$, where $\Pi$ is a permutation), then $U(p) \rho_{sb} U(p) = U \rho_{sb} U^{\dagger}$ for all $p$, and hence $\rho_{sb}' = U \rho_{sb} U^{\dagger}$ exactly.

Alternatively, suppose that the initial momentum distribution of the weight $\mu(p)$ satisfies 
\begin{align} \label{eq::momdist} 
\int_{-\epsilon}^{\epsilon} dp \; \mu(p) \geq 1-\delta.
\end{align} 
for small $\epsilon$ and $\delta$. 

The final state of the system and bath $\rho_{sb}'$ is given by \eqref{eq:finalstateexplicit}. The trace-distance between this state and the desired final state $U \rho U^{\dagger}$ can then be bounded as follows. 
\begin{align} 
D(\rho_{sb}',U \rho_{sb} U^{\dagger}) &=  \frac{1}{2}  \| \rho_{sb}' - U \rho_{sb} U^{\dagger} \|_1 \nonumber \\
&\leq \frac{1}{2}\int dp \mu(p)   \| U(p) \rho_{sb} U(p)^{\dagger} - U \rho_{sb} U^{\dagger} \|_1 \nonumber \\
&\leq \delta \!+\! \frac{1}{2} \int_{-\epsilon}^{\epsilon} \!\!dp \mu(p)   \| U(p) \rho_{sb} U(p)^{\dagger}\!\! -\! U \rho_{sb} U^{\dagger} \|_1  \nonumber \\
& \leq \delta+ \frac{1}{2} \int_{-\epsilon}^{\epsilon} \mu(p) ( 4 |p| \|H_{sb}\| + O(p^2) ) \nonumber \\
&\leq  \delta + O(\epsilon),
\end{align} 
where the third step can be seen by expanding \eqref{eq:explicitU(p)} as a power series in $p$, \begin{align}
U(p) = U -i p (U H_{sb} - H_{sb} U) +O(p^2)
\end{align} 
and using $\| H \rho \|_1 \leq \|H\|$ where $\rho$ is a density operator. 
Hence by using a weight state with sufficiently small $\epsilon$ and $\delta$ we can make the final state of the system and bath arbitrarily close to the desired final state. As the system and bath have a bounded spectrum, this means their final expected energy must be arbitrarily close to the desired value. As energy is conserved overall, this means that the weight must also be raised by and amount  (i.e. work must be done) that is arbitrarily close to the desired value from the implicit battery case. 

As a concrete example, consider a weight in a pure `top-hat' wavefunction of width $L$ 
\begin{equation} \ket{\psi_L}=\frac{1}{\sqrt{2L}}\int_L^Ldx \ket{x}, \end{equation} 
with $\rho_w=\ket{\psi_L}\bra{\psi_L}$. For this state 
\begin{equation} 
\mu(p) = \frac{\hbar}{\pi L} \frac{\sin^2\left( \frac{p L}{\hbar} \right)}{p^2}  
\end{equation} 
Taking $\epsilon = \frac{c}{\sqrt{L}}$, for some constant $c$, we find that 
\begin{align} 
\int_{-\epsilon}^{\epsilon} dp \, \mu(p) &\geq 1-2\int_\epsilon^{\infty} dp\,\mu(p)  \nonumber \\
&\geq 1-\frac{2 \hbar}{\pi L} \int_{\frac{c}{\sqrt{L}}}^{\infty} dp\, \frac{1}{p^2} \nonumber \\
& \geq 1- \frac{2 \hbar}{\pi  c \sqrt{L}}
\end{align} 
hence we have that both $\epsilon\rightarrow 0$ and $\delta \rightarrow 0$ as $L \rightarrow \infty$.

\section{An alternative quantifier of work for channels}\label{app::alternatequantifier}

In the main paper we require that measurements and channels must draw any energy they require to operate from an external battery. This means that they are treated in the same way as the unitary transformations allowed by the framework.

An alternative possibility is to consider the work benefit of a device which has its own internal power supply which we do not include in our accounting (or a device on which we simply do not impose energy conservation). In this case, we take the work benefit of applying the device to be zero in all cases ($W^{\text{apply}}=0$). In this appendix, we explore how our results change in this case. 

In the case of a channel $C$, all work benefit is then obtained during the process of resetting the final state $\sigma_s$ to its initial state $\rho_s$, and is simply dependent on the free energy change in this process. Following a similar proof to before with $W^{\text{apply}}_C=0$, and noting that $E(\sigma_a) = E(\rho_a)$,  we obtain 
\begin{align} 
W^{\mathrm{total}}_{C,P} \leq F(C[\rho_t]) - F(\rho_t)
\end{align} 
and 
\begin{align} 
W^{\mathrm{total}}_{C} = \max_{\rho_t} \left( F(C[\rho_t]) - F(\rho_t) \right). 
\end{align} 
Note that even if the channel is unitary, $W^{\mathrm{total}}_{C}$ will generally be non-zero, because of energy differences the channel can induce. When considering consistency with the second law, we must now consider both entropy changes and energy changes inside the device.

In the main text, it was shown that  channels are useless for work extraction if and only if they are unital. Can we formulate a similar result  when we do not associate a work cost for the channel? It turns out that we can, only now the useless class of channels are the Gibbs preserving maps. A channel is Gibbs preserving if it preserves the thermal state, $C[\tau]=\tau$, for $\tau=e^{-\frac{ H}{T}}/Z$ were $Z=\tr{e^{-\frac{ H}{T}}}$. Let us show this now. 

Consider the relative entropy between states $D(\rho||\tau)=\texttt{Tr}[\rho \ln \rho] - \texttt{Tr}[\rho \ln \tau]$. We have by the data processing inequality that $D(C[\rho]||C[\tau])\leq D(\rho||\tau)$. Let us first assume that the map $C$ is Gibbs preserving. Direct substitution then gives us 
\begin{align}
&   \tr{C[\rho] ( \ln C[\rho] - \ln C[\tau])} \leq \tr{\rho (\ln \rho - \ln \tau)}\nonumber\\
&\implies   \tr{C[\rho] (\ln C[\rho] - \ln \tau)} \leq \tr{\rho (\ln \rho -  \ln \tau)}\nonumber \\
&\implies S(\rho)  -S(C[\rho]) \leq \tr{(C[\rho]-\rho) \ln \tau}\nonumber\\
&\implies S(\rho)  -S(C[\rho]) \leq  \tr{(C[\rho]-\rho) \left({\scriptstyle -\frac{H}{T}}- \ln Z\right)} \nonumber\\
&\implies S(\rho)  -S(C[\rho]) \leq  \frac{1}{T}\left( E(\rho)-E(C[\rho]) \right) \nonumber\\
& \implies F(C[\rho]) \leq F(\rho)
\end{align}
Which shows that if $C$ is a Gibbs preserving map, it cannot increase the free energy, and as such is useless for work extraction. Let's show the other direction as well. Suppose a map is useless for work extraction, i.e. it cannot increase the free energy. Then we have that $F(C[\tau])\leq F(\tau)$. The thermal Gibbs state is the state which is the unique minimiser for the free energy. This means that this inequality is an equality, and it follows that $C[\tau]=\tau$, and hence the map is Gibbs preserving. 

For measurements, taking $W^{\text{apply}}_{\{C_i\}}=0$, we similarly have 
\begin{align}
 W^{\mathrm{total}}_{\{C_i\}} = \max_{\rho_t} T \left(  \sum_i p_i F(\sigma_{t,i}) - F(\rho_t) \right).
\end{align}
In this case, the conditional work benefit when obtaining result $i$ is simply
\begin{align} 
 W^{\mathrm{total}}_{\{C_i\}|i, \rho_s} =  T \left(   F(\sigma_{s,i}) - F(\rho_s) \right).
 \end{align} 

Finally,  we still obtain an unbounded work benefit for any (non-trivial) post-selected measurement. Applying the measurement to the same state as in \eqref{eq::post-selectedstate} and following a similar proof to that in section \ref{sec::postselection}, we obtain 
\begin{align} 
 W^{\mathrm{total}}_{\{C_i\}|\text{succ}, \rho_s} &\geq  W^{\mathrm{total}}_{\{C_{\text{succ}}, C_{\text{fail}}\}|\text{succ}, \rho_s} \nonumber \\
 &\geq F(\sigma_{s,\text{succ}}) - F(\rho_s) \nonumber \\
 & \geq E_{\text{min}} - E_{\text{max}} \nonumber \\ &\quad + T \left( S(\rho_s)- S(\sigma_{s, \text{succ}}) \right) \nonumber \\
 & \geq E_{\text{min}} - E_{\text{max}} \nonumber \\ &\quad + T \left(\left(\frac{1}{2}-q \right) \ln d_a - \ln d_t \right).
\end{align} 
which can be made arbitrarily large by increasing $d_a$.

\section{A unital non-catalytic channel} \label{app::unital}

For a two dimensional target system it is known that every unital channel can be expressed as a convex combination of unitary transformations \cite{Ritter2005}. These are catalytic, as they could be implemented via a controlled unitary from a mixed ancilla onto the system. However, in higher dimension this is not always the case. Consider the 
 Werner-Holevo channel \cite{Werner2002} for $d=3$:
\begin{equation}\label{eq:WHchanappendix}
    \rho \rightarrow C^{\text{WH}}(\rho)=\frac{1}{2}(\tr{\rho}\identity-\rho^T), 
\end{equation}
where $\rho^T$ corresponds to the transpose of $\rho$ in a particular basis. This channel can easily be seen to be unital and  non-unitary \footnote{To see that it is non-unitary, consider applying it to a projector in the basis in which the transpose is taken, and note that the eigenvalues are changed.}. A key property of this channel is that it is extremal in the space of channels, and therefore cannot be expressed as a mixture of different channels. 

We will now construct a proof by contradiction to show that $C^{\text{WH}}$ is not a catalytic channel. Suppose that there was a catalytic implementation of $C^{\text{WH}}$ 
\begin{align} 
C^{\text{WH}}[\rho_t] &= \trs{W \rho_t \otimes \rho_z W^{\dagger}}{z} \\
\rho_z &= \trs{W \rho_t \otimes \rho_z W^{\dagger}}{t}
\end{align}
As $\rho_z$ remains unchanged by the evolution, the unitary $W$ cannot take states within the support of $\rho_z$ outside of this support. Hence without loss of generality we can restrict $Z$ to the support of $\rho_z$. For any state $\ket{\phi}_z$, there then exists a sufficiently small $\epsilon$ such that $\rho_z = \epsilon \proj{\phi} + (1-\epsilon) \sigma_z$, where $\sigma_z$ is a valid state. 

This implies that  
\begin{align} 
C^{\text{WH}}[\rho_t] = \epsilon C_{\phi} [\rho_t] + (1-\epsilon) C_{\sigma} [\rho_t]
\end{align} 
where 
\begin{align} 
 C_{\phi} [\rho_t]&= \trs{W \rho_t \otimes \proj{\phi}  W^{\dagger}}{z} \\
  C_{\sigma} [\rho_t]&= \trs{W \rho_t \otimes \sigma_z  W^{\dagger}}{z}
 \end{align}
are both valid channels. As $C^{\text{WH}}$ is extremal, it must be the case that 
$C_{\phi} [\rho_t]  =C^{\text{WH}}[\rho_t]$ for all $\ket{\phi}$. It has been shown \cite{beckman2001} that this independence on the state of the ancilla implies that $W$ is a product unitary. I.e. $W = W_t \otimes W_z$. This would imply that $C^{\text{WH}}$ is the unitary channel $C^{\text{WH}} [\rho_t] = W_t \rho_t W_t^{\dagger}$. However, since $C^{\text{WH}}$ is non-unitary we obtain a contradiction. Therefore  $C^{\text{WH}}$ is not a catalytic channel.

\section{The implications of having access to larger system sizes}
Here we show that in the case of measurements, the total work benefit of the measurement depends only on the target system to which it is applied, and not any ancilla in the system. For example, it does not matter if the target system is maximally mixed, or half of a maximally entangled pure state, and does not increase when the ancilla is allowed to grow in size.      

Recall that the system $S$ is composed of the target system $T$ and the ancilla $A$. Given the results in the main paper, we wish to show that 
\begin{align}\label{eq::largersystems}
S(\rho_{s}) - \sum_{i} p_i S(\sigma_{s,i}) \, \leq \, S(\rho_{t}) - \sum_{i} p_i S(\sigma_{t,i}) 
\end{align}
where $\sigma_{s,i}$ is the post measurement state of the system on obtaining result $i$, and $\sigma_{t,i}$ is the post measurement state of the target alone. We can always represent the effect of the measurement by Kraus operators, such that 
$\sigma_{s,i} = \frac{1}{p_i}\sum_{j} \left(\mathcal{I} \otimes K_{ij} \right) \rho_{s} \left(\mathcal{I} \otimes K_{ij}^{\dagger} \right)$ with $p_i = \sum_{j} \tr{ K_{ij}^{\dagger} K_{ij} \rho_t}$ and $\sigma_{t,i} = \trs{\sigma_{s,i}}{a}$ . 

To prove the desired relation, we introduce a third quantum system $X$, which stores the measurement result (note that this will typically be a system inside the physical measuring device, but here we introduce it only as a mathematical construction). 

 Recall that the mutual information is non-increasing under action of a local channel
\begin{align}
    I(A:B) \geq I(A:C[B])
\end{align}
which implies that 
\begin{align}\label{eq::entropysystems}
    S(AB)-S(\mathcal{I}\otimes C[AB]) \leq S(B)-S(C[B]).
\end{align}

In our particular case, we will take $A$ to be the ancilla, and $B$ to be composed of the target and measurement result, so $\mathcal{H}_B=\mathcal{H}_T \otimes \mathcal{H}_X$. As far as measurements are concerned, we are interested in channels of the form 
\begin{equation}
    C[\rho_{tx}]=\sum_{ij} \left( K_{ij}\otimes\Gamma_i \right) \rho_{tx} \left( K_{ij}^{\dagger}\otimes\Gamma_i^{\dagger} \right) 
\end{equation}
where the $\Gamma_i$ here are discrete unitary shift operators which act to increment the measurement result  by $i$, $\Gamma_i\ket{0}_X=\ket{i}_X$. Taking $\rho_{sx} = \rho_s \otimes \proj{0}$ and substituting all into equation \eqref{eq::entropysystems} we find that
\begin{align}
    S(\rho_s \otimes \proj{0})-S(\sum_{ij} \left(\mathcal{I} \otimes K_{ij} \right) \rho_{s} \left(\mathcal{I} \otimes K_{ij}^{\dagger} \right) \otimes \ket{i}\bra{i}  )  \nonumber\\
    \leq S(\rho_t \otimes \proj{0}) - S(\sum_{ij} K_{ij} \rho_t K_{ij}^{\dagger} \otimes \ket{i}\bra{i})
\end{align} 
and hence 
\begin{align}     
    S(\rho_s)&-S(\sum_{i} p_i \sigma_{s,i}  \otimes \ket{i}\bra{i})  \nonumber\\
    &\leq S(\rho_t) - S(\sum_{i} p_i\sigma_{t,i} \otimes \ket{i}\bra{i})
\end{align}
We now use the fact that for a classically correlated state $S(\sum_{i} p_i \sigma_i  \otimes \ket{i}\bra{i})=\sum_{i}p_{i}S(\sigma_{i})+H(p_i)$, where $H(p_i)$ denotes the Shannon entropy of the distribution specified by $p_i$. We therefore find
\begin{align}
     S(\rho_s)& -\sum_{i}p_i S( \sigma_{s,i}) + H(p_i) \nonumber \\
     &\leq 
    S(\rho_t) - \sum_{i}p_i S( \sigma_{t,i}) +H(p_i)
\end{align}
which proves the result \eqref{eq::largersystems}.

\section{Evaluating the conditional work benefit of measurement}\label{app::workevaluations}

In this appendix, we calculate the conditional work benefit $W_{\{C_i\}|i}^{\text{apply}}$ of applying a measurement to the target system and obtaining result $i$, using explicit batteries. 

Recall that the system $s$ is composed of an  ancilla $a$ and the target system $t$, there is also an additional ancilla inside the measuring device $z$ (which stores the measurement result, but may also have other degrees of freedom), and the `weight' $w$ which stores/provides any change in work. In the implicit battery formalism, the  measurement is carried out by implementing a unitary transformation on $t$ and $z$, and then performing a projective measurement on $z$. 
\begin{equation} 
C_i[\rho_t] = \trs{V \rho_t \otimes \rho_{z} V^{\dagger} (I \otimes \Pi_i}{z}, 
\end{equation} 
where $V$ is a unitary, and $\Pi_i$ is a projector onto the outcome space corresponding to result $i$. Expanding $V$ in an energy basis as \begin{equation}
    V=\sum_{a b c d}\alpha_{a b}^{c d}\ket{a}\bra{b} \otimes \ket{c} \bra {d},
\end{equation}
we can then extend this to the explicit battery formalism by considering  
\begin{equation}
     {\tilde{V}}=\sum_{a b c d}\alpha_{a b}^{c d}\ket{a}\bra{b} \otimes \ket{c} \bra {d} \otimes \Gamma_{E_b-E_a}.
\end{equation}
where $\Gamma_E$ is a translation operator for the weight. Including the weight gives   
\begin{equation} \label{eq::chanimplicit2}
C_i[\rho_t] \simeq \trs{\tilde{V} \rho_t \otimes \rho_{z} \otimes \rho_w \tilde{V}^{\dagger} (I \otimes \Pi_i \otimes I )}{zw}, 
\end{equation} 
where $\rho_w$ is the initial state of the weight. The approximation here is due to the width of the weight state. Considering a `top-hat' wavefunction for the weight, $\rho_w = \proj{\psi_L}$ where
\begin{equation} \ket{\psi_L}=\frac{1}{\sqrt{2L}}\int_{-L}^Ldx \ket{x},
\end{equation}
we obtain corrections of $O(\frac{1}{L})$ which can be made as small as desired by taking $L$ sufficiently large. 

The conditional work benefit of applying the measurement is given by the average of $x$ in the final weight state given that result $i$ was obtained, minus the initial average of $x$. As the latter is zero  ($\bra{\psi_L} \hat{x} \ket{\psi_L}=0$), we have 
\begin{equation}
     W^{\mathrm{apply}}_{\{C_i\}|i}=\frac{1}{p_i}\tr{(\tilde{V} \rho_t \otimes \rho_z \otimes \tau_w \tilde{V}^{\dagger})( \mathbf{I} \otimes \Pi_i \otimes \hat{x})]}
\end{equation}
Substituting in we find that 
\begin{align}\label{eq::startofworkcal}
    W^{\mathrm{apply}}_{\{C_i\}|i}=& \frac{1}{p_i} \text{Tr} \bigg(\sum_{a b c d}\alpha_{a b}^{c d}\ket{a}\bra{b} \otimes \ket{c} \bra {d} \otimes \Gamma_{E_b-E_a}\nonumber\\&(\rho_t \otimes \rho_z \otimes \rho_w)\sum_{e f g h}\alpha_{e f}^{g h *}\ket{f}\bra{e} \otimes \ket{h} \bra {g} \otimes \Gamma_{E_e-E_f}^{\dagger}\nonumber\\&\qquad( \mathbf{I} \otimes \Pi_i \otimes \hat{x})\bigg)\nonumber\\
    &=\frac{1}{p_i}\sum_{a b c d e f g h}\alpha_{a b}^{c d}\alpha_{e f}^{g h *} \textrm{Tr} \bigg(\ket{a}\bra{b} \rho_t\ket{f}\bra{e}\otimes \nonumber\\& \ket{c} \bra {d} \rho_z \ket{h} \bra {g}\Pi_i \otimes \Gamma_{E_b-E_a} \rho_w \Gamma_{E_e-E_f}^{\dagger}\hat{x} \bigg).
\end{align}
Next, we will demonstrate how to evaluate the trace over the weight system. Consider the general case 
$
    \texttt{Tr}[\Gamma_A \rho_w \Gamma_B^{\dagger} \hat{x}]
$
where $A$ and $B$ are constants. Using the cyclic symmetry of the trace we have 
\begin{align} \label{eq::traceweight} 
  \texttt{Tr}[\Gamma_A \rho_w \Gamma_B^{\dagger} \hat{x}] &= \int_{-\infty}^{\infty} dx \,x\, \bra{x}  \Gamma_A \rho_w \Gamma_B^{\dagger}\ket{x} \nonumber\\
     &=\int_{-\infty}^{\infty} dx \,x\,\bra{x-A} \rho_w \ket{x-B} \nonumber\\
    &= \int_{-\infty}^{\infty} dx \,x\; \psi_{L}(x-A)\psi_{L}(x-B)\nonumber \\
    & = \frac{1}{2L} \int_{\max(A,B) -L}^{\min(A,B)+L} dx \, x \nonumber \\
    & = \frac{1}{4L}\left( (\min(A,B)\!+\!L)^2 -(\max(A,B)\!-\!L)^2 \right)  \nonumber\\
    & = \frac{1}{2}(A+B) - \frac{|A^2-B^2|}{4L}.
\end{align}
Considering very large $L$ for which we can neglect corrections of $ O(\frac{1}{L})$, we get $\frac{1}{2}(A+B)$ . Setting $A=E_f-E_e$ and $B=E_b-E_a$ and substituting the result into \eqref{eq::startofworkcal} we find that $ W^{\mathrm{apply}}_{\{C_i\}|i}$ can be given as
\begin{align}
    W^{\mathrm{apply}}_{\{C_i\}|i}&=\frac{1}{p_i}\sum_{a b c d e f g h}\alpha_{a b}^{c d}\alpha_{e f}^{g h *} \frac{(E_f-E_e)+(E_b-E_a)}{2} \nonumber \\ & \qquad \textrm{Tr} \bigg(\ket{a}\bra{b} \rho_t\ket{f}\bra{e}\otimes \ket{c} \bra {d} \rho_z \ket{h} \bra {g} \bigg) \nonumber \\
    &=\frac{1}{p_i}\sum_{a b c d e f g h}\alpha_{a b}^{c d}\alpha_{e f}^{g h *} \nonumber \\ &\textrm{Tr} \bigg(\ket{a}\bra{b} \frac{(H_t\rho_t+\rho_t H_t)}{2}\ket{f}\bra{e}\otimes  \ket{c} \bra {d} \rho_z \ket{h} \bra {g}  \nonumber \\
    & \qquad - H_t \ket{a}\bra{b} \rho_t \ket{f}\bra{e}\otimes \ket{c} \bra {d} \rho_z \ket{h} \bra {g} \bigg) \nonumber \\
    &= \frac{1}{p_i} \texttt{Tr} \bigg( V\frac{(H_t\rho_t+\rho_t H_t)}{2}\otimes \rho_z V^{\dagger} - H_t V\rho_t \otimes \rho_z V^{\dagger} \bigg) \nonumber \\
    &=\frac{1}{p_i}\texttt{Tr} [{C}_i( \frac{H_t\rho_t+\rho_t H_t}{2}) - H_t {C}_i(\rho_t)]
\end{align}
from which we recover \eqref{eq::thrm3}.

\section{Weight state dependence}\label{app::otherweights}
In order that a device correctly implements a channel or measurement when extended to an explicit weight state, it is only necessary that the weight state has a momentum distribution sharply peaked about zero (as in \eqref{eq::momdist}). This is also the case for the total work benefit of channels and measurements. However, when considering conditional or post-selected work benefits we show here that the result has further dependence on the initial state of the weight. 

So far, we have used a broad top-hat function $\ket{\psi_L}=\frac{1}{2L}\int_{-L}^{L}\ket{x}dx$ for the initial state of the weight. Let us now consider how the calculation in appendix \ref{app::workevaluations} changes for different states of the weight.  

The dependence on the initial weight state comes when calculating $ \tr{\Gamma_A \rho_w \Gamma_B^{\dagger} \hat{x}} $ in  \eqref{eq::traceweight}. We first show that any real symmetric initial wavefunction for the weight $\phi_L(x)$, which satisfies 
\begin{equation} 
\int_{-\infty}^{\infty} dx \, \phi_L(x-C)  \phi_L(x+C) = 1 + O\left(\frac{1}{L}\right)
\end{equation} 
in the large $L$ limit will give the same results as the top-hat wavefunction. This is because 
\begin{align} 
\tr{\Gamma_A \rho_w \Gamma_B^{\dagger} \hat{x}} &= \int_{-\infty}^{\infty} dx \, x\phi_L(x-A)  \phi_L(x-B) \nonumber \\
&= \int_{-\infty}^{\infty}\!\! dx \, \left(x+\scriptstyle{\frac{A+B}{2}}\right) \phi_L\left(x-\scriptstyle{\frac{A-B}{2}}\right)  \phi_L\left(x+\scriptstyle{\frac{A-B}{2}}\right)   \nonumber \\
&= \frac{A+B}{2} \int_{-\infty}^{\infty} dx\, \phi_L\left(x-\scriptstyle{\frac{A-B}{2}}\right)  \phi_L\left(x+\scriptstyle{\frac{A-B}{2}}\right) \nonumber \\
&= \frac{A+B}{2} + O\left(\frac{1}{L}\right),
\end{align} 
where in the third line we have used the symmetry of $\phi_L(x)$. This means that other common weight states such as Gaussians will also work. 

However, let us now consider an asymmetric `triangular' weight state 
\begin{equation}
    \ket{\tau_w} = \sqrt{\frac{3}{L^3}}\int_{-\frac{3L}{4}}^{\frac{L}{4}}(x+{\scriptstyle \frac{3L}{4}})\ket{x}dx, 
\end{equation}
where the constants are chosen so that the state is normalised, and $\bra{\tau_L} \hat{x} \ket{\tau_L}=0$. Note that this state still has a narrow momentum distribution about $p=0$ in the limit of large $L$. 

Taking $\rho_w = \proj{\tau_L}$ we have
\begin{align} 
&\tr{\Gamma_A \rho_w \Gamma_B^{\dagger} \hat{x}} \nonumber \\
\quad&= \int_{-\infty}^{\infty} dx \, x\tau_L(x-A)  \tau_L(x-B) \nonumber \\
\quad&= \frac{3}{L^3} \int_{\max(A,B)-\frac{3L}{4}}^{\min(A,B) + \frac{L}{4}} dx \, x (x+{\scriptstyle\frac{3L}{4}}-A)(x+{\scriptstyle\frac{3L}{4}}-B)\nonumber \\
\quad&=\frac{A+B}{8} +\frac{3 \min(A,B)}{4} +  O\left(\frac{1}{L}\right) \nonumber \\
\quad &=\frac{A+B}{2} -\frac{3 |A-B|}{8} +  O\left(\frac{1}{L}\right) \nonumber \\
\end{align} 
Note that the first term is the same as we got for the `top-hat' weight state. To calculate the effect of the second term, it is helpful to express the measurement using a Kraus decomposition. 
\begin{align} 
C_i[\rho_t] = \sum_j K_{ij} \rho_t K_{ij}^{\dagger}.  
\end{align}
Extending this to incorporate an explicit weight, we define 
\begin{equation}
\tilde{K}_{ij}=\sum_{a,b}(K_{ij})_{ab} \ket{a}\bra{b} \otimes \Gamma_{E_b-E_a}. 
\end{equation}
Then
\begin{align}
     W^{\mathrm{apply}}_{\{C_i\}|i}&= \sum_j \frac{1}{p_i}\tr{(\tilde{K}_{ij} \rho_t  \otimes \tau_w \tilde{K}_{ij}^{\dagger})( \mathbf{I}\otimes \hat{x})]} \nonumber \\
     &=\frac{1}{p_i}\sum_{jabcd} (K_{ij})_{ab} (K^{\dagger}_{ij})_{cd} \nonumber \\ 
     &\qquad \tr{\ket{a}\bra{b} \rho_t \ket{c}\bra{d} \otimes \Gamma_{E_b-E_a} \rho_w \Gamma^{\dagger}_{E_c-E_d}\hat{x}} \nonumber \\
     &= \frac{1}{p_i}\sum_{jabcd} (K_{ij})_{ab} (K^{\dagger}_{ij})_{cd} \nonumber \\ 
     &\qquad \text{Tr} \bigg(\ket{a}\bra{b} \frac{H_t\rho_t+\rho_t H_t}{2} \ket{c}\bra{d} \nonumber \\ 
     &\qquad \qquad -  H_t \ket{a}\bra{b} \rho_t \ket{c}\bra{d} \nonumber \\
     &\qquad \qquad -\frac{3}{8} |E_b-E_a -E_c+E_d| \ket{a}\bra{b} \rho_t \ket{c}\bra{d} \bigg) \nonumber \\
     &= \tr{C_i\left[\frac{H_t \rho_t +\rho_t H_t}{2}\right]}-\tr{H_t C_i[\rho_t]}\nonumber \\
     &\qquad - \frac{3}{8p_i}\sum_{jabc} (K^{\dagger}_{ij})_{ca}(K_{ij})_{ab} |E_b-E_c|\bra{b}\rho_t\ket{c} \nonumber\\
     &= \tr{C_i\left[\frac{H_t \rho_t +\rho_t H_t}{2}\right]}-\tr{H_t C_i[\rho_t]}\nonumber \\
     &\qquad - \frac{3}{8 p_i}\sum_{bc} (M_i)_{cb} |E_b-E_c|\bra{b}\rho_t\ket{c} \nonumber
\end{align}
where $M_i = \sum_j K_{ij}^{\dagger} K_{ij}$ is the POVM element corresponding to result $i$ (which satisfies $p_i = \tr{M_i \rho_t}$), which is independent of the  choice of Kraus decomposition. 
\vspace{1cm} 

Note that the extra term above compared to our previous result is dependant on the off-diagonal elements of $\rho$ and the measurement operator $M$ in its energy eigenbasis. This state of the weight produces a different energy shift  to the top-hat state of the weight when the energy and state contain energy coherences, even when $L\rightarrow \infty$. This suggests that the conditional work benefit of applying a measurement is not operationally well defined without specifying the weight state in detail. 
It is not clear how large the set of weight states is which lead to an additional work benefit term compared the top-hat weight state. Perhaps there are only a few `bad' weight states which lead to this (e.g. those containing both spatial non-continuity and asymmetry), in which case the result of theorem \ref{thm3} could reasonably be taken as standard. This is an interesting avenue for further investigation.

\end{document}